\documentclass[10pt]{article}
\usepackage{a4wide}                      
\usepackage{epsf}                        
\usepackage{latexsym}                    
\usepackage{amsfonts}                    
\usepackage{amssymb}                     

\newcommand{\sect}[1]{\setcounter{equation}{0}\section{#1}}
\renewcommand{\theequation}{\arabic{section}.\arabic{equation}}

\def\be{\begin{equation}}
\def\ee{\end{equation}}
\def\bea{\begin{eqnarray}}
\def\eea{\end{eqnarray}}
\def\nn{\nonumber \\ [.2cm]}

\def\hsp#1{\hspace{#1}}

\def\part{\partial}
\def\tfrac#1#2{{\textstyle{\frac{#1}{#2}}}}

\def\x{\times}

\def\Tr{\mbox{Tr}}

\def\incl{{\mbox{\footnotesize i}}}
\def\ii{{\incl_X \incl_X}}
\def\cA{{\cal A}}
\def\cL{{\cal L}}
\def\cF{{\cal F}}
\def\tC{{\tilde C}}

\def\ux{{\underline x}}

\def\hmu{{\hat \mu}}
\def\hnu{{\hat \nu}}
\def\hrho{{\hat \rho}}
\def\mn{{\mu\nu}}

\def\hmn{{\hmu\hnu}}

\def\makeatletter{\catcode`\@=11}
\makeatletter
\def\mathbox#1{\hbox{$\m@th#1$}}%
\def\math@ccstyles#1#2#3#4#5#6#7{{\leavevmode
      \setbox0\mathbox{#6#7}%
      \setbox2\mathbox{#4#5}%
      \dimen@ #3%
      \baselineskip\z@\lineskiplimit#1\lineskip\z@
      \vbox{\ialign{##\crcr
             \hfil \kern #2\box2 \hfil\crcr
             \noalign{\kern\dimen@}%
             \hfil\box0\hfil\crcr}}}}
\def\mathaccstyles{\math@ccstyles\maxdimen}
\def\maththroughstyles{\math@ccstyles{-\maxdimen}}
\def\unity%
 {\maththroughstyles{.45\ht0}\z@\displaystyle {\mathchar"006C}\displaystyle 1}
\begin{document}

\rightline{KUL-TF/03-35}
\rightline{FFUOV-03/16}
\rightline{hep-th/0312264}
\rightline{January 2004}
\vspace{1truecm}

\centerline{\LARGE \bf The gauge invariance of the non-Abelian  }
\vspace{.5truecm}
\centerline{\LARGE \bf Chern-Simons action for D-branes revisited}
\vspace{1.3truecm}

\centerline{
    { \bf J. Adam${}^{a,}$}\footnote{E-mail address:
                                  {\tt joke.adam@fys.kuleuven.ac.be}},
    { \bf J. Gheerardyn${}^{a,}$}\footnote{E-mail address:
                                  {\tt jos.gheerardyn@fys.kuleuven.ac.be}},
    { \bf B. Janssen${}^{a,}$}\footnote{E-mail address:
                                  {\tt bert.janssen@fys.kuleuven.ac.be} }
    {\bf and}
    { \bf Y. Lozano${}^{b,}$}\footnote{E-mail address:
                                  {\tt yolanda@string1.ciencias.uniovi.es}}
                                                            }

\vspace{.4cm}
\centerline{{\it ${}^a$ Instituut voor Theoretische Fysica, K.U. Leuven}}
\centerline{{\it Celestijnenlaan 200D,  B-3001 Leuven, Belgium}}

\vspace{.4cm}
\centerline{{\it ${}^b$Departamento de F{\'\i}sica,  Universidad de Oviedo,}}
\centerline{{\it Avda.~Calvo Sotelo 18, 33007 Oviedo, Spain}}

\vspace{2truecm}

\centerline{\bf ABSTRACT}
\vspace{.5truecm}

\noindent
We present an elegant method to prove the invariance of the Chern-Simons part of the
non-Abelian action for $N$ coinciding D-branes under the R-R and NS-NS gauge transformations,
by carefully defining what is meant by a background gauge transformation in the non-Abelian
world volume action.  We study as well the invariance under massive gauge transformations of
the massive Type~IIA supergravity and show that no massive dielectric couplings are
necessary to achieve this invariance. We show that this result is consistent with (massive)
T-duality from the non-Abelian action for $N$ D9-branes.

\newpage
\sect{Introduction}

It is well known by now that the physics of a set of $N$ coincident D$p$-branes can be
very different from the physics of $N$ parallel but separated D$p$-branes. The latter is
described by a $(p+1)$-dimensional Abelian world volume action, with as bosonic field content
$N$ $U(1)$ vector fields $V_a^I$, with $a$ the world volume index, and
$N$ times $(9-p)$ scalars $X^{iI}$, that can be
arranged in $(9-p)$ $N\x N$ diagonal matrices. The $U(1)$ vectors are of course the Born-Infeld
vectors living on each brane, while a scalar $X^{iI}$ represents the position of the $I$-th
brane in the transversal direction $x^i$. As the separation between the different D-branes
decreases, the open strings stretched between two distinct branes grow shorter and lighter,
so that in the limit where all
D-branes coincide new massless states are generated, and new physics appears.

Witten showed \cite{Witten} that these new massless states are arranged such that the $U(1)^N$
gauge symmetry of the system of D-branes is enhanced to a full non-Abelian $U(N)$ gauge symmetry.
The $N$ Born Infeld vectors form a single $U(N)$ Yang-Mills vector $V_a$ and the
transverse scalars, arranged in $N \x N$ matrices $X^i$, become non-Abelian matrices transforming
in the adjoint representation of $U(N)$. The $I$-th eigenvalue of the matrix $X^i$ has still
the interpretation of the position of the $I$-th D-brane in the direction $x^i$, but since in
the $U(N)$ case not all matrices are simultaneously diagonalisable, the branes are no longer
fully localisable in all transverse directions. Therefore, the geometry of the transverse space
described in terms of the matrix-valued coordinates $X^i$
becomes that of a ``fuzzy surface''.

The new physics associated to these extra massless string states has to be encoded in the world
volume effective action describing the system of coincident branes. This action
should now be written in terms of the matrix valued fields $V_a$
and $X^i$. Determining the exact form of the
Born-Infeld action is a highly non-trivial problem, to
which the solution is still not clear (see for instance \cite{Tseytlin2}).
A lot of progress has been made however over the last few years in the understanding
of the structure of the non-Abelian Chern-Simons (or Wess-Zumino) action.

The first generalisation of the Chern-Simons term
to the $U(N)$ case was proposed in \cite{Douglas}:
\bea
S_{{\rm D}p}\ =\ T_p \int P[C] \ \Tr \{ e^\cF \}
            \ = \ T_p \int \sum_n P[C_{p-2n+1}]\  \Tr \{ \cF^n \}.
\label{GHTac}
\eea
Here the trace is taken over the Yang-Mills indices of the $N$-dimensional representation
of $U(N)$ and  $P[\Omega]$ denotes the pullback of the background field $\Omega$ to the
world volume of the D-brane. The world volume field $\cF$ is given by $\cF = F + P[B]$, where
$F_{ab} = 2 \part_{[a} V_{b]} + i [V_a, V_b]$ is the non-Abelian field strength of the
Born-Infeld vector and $B$ the NS-NS two-form.

The invariance of this action under the gauge transformations of the background
NS-NS and R-R fields was further investigated in \cite{GHT}, where it was shown
as well that in order to be invariant under the massive gauge transformations of massive
Type IIA supergravity \cite{Romans, BRGPT}, extra $m$-dependent terms were needed in the
action. These extra terms were also obtained from the (massive) T-duality relations
\cite{BRGPT, BR} between the different D-brane actions, generalising to the non-Abelian
case the Abelian calculation of \cite{BR}.

Nowadays we know, however, that the Chern-Simons action
for coincident D-branes presented in \cite{GHT} is not the complete story.
On the one hand, in the non-Abelian case the background fields in (\ref{GHTac}) must be
functionals of the matrix-valued coordinates $X^i$ \cite{Douglas2}.
Explicit calculations of string scattering amplitudes \cite{GM} suggest that
this dependence is given by a non-Abelian Taylor expansion
\be
C_\mn (x^a, X^i) = \sum_n \frac{1}{n!} \part_{k_1} ... \part_{k_n}
                     C_\mn (x^a, x^i){\mid}_{x^i = 0} \
                      X^{k_1} ... \ X^{k_n}.
\label{taylorexp}
\ee
On the other hand, in order to have invariance under $U(N)$ gauge transformations the
pullbacks of the background fields into the world volume have
to be defined in terms of $U(N)$ covariant derivatives
$D_a X^\mu = \part_a X^\mu + i[V_a, X^\mu]$,
rather than partial derivatives \cite{Dorn, Hull}.
For instance\footnote{From now on instead of working in the static gauge we will write everything
in a ``diffeomorphism invariant'' way (see however \cite{DS, Hassan, DSW} and our
comments in the conclusion), with the understanding that $U(N)$ covariant derivatives reduce to
ordinary ones for $X^\mu$ laying in the world volume of the D-branes.},
\be
P[C_2] = C_\mn \ D_{[a}X^\mu D_{b]}X^\nu.
\label{pullback}
\ee
This, together with the symmetrised trace prescription of Tseytlin
\cite{Tseytlin}\footnote{Although for the non-Abelian Born-Infeld part of the action, the
symmetrised trace prescription receives corrections at order $(\alpha')^3$
\cite{HT, Bain, DST, STT}, there is no reason to think that it would not be valid for the
Chern-Simons part of the action.},
that we will denote by curly brackets $\{..\}$, assures the
invariance of the action under $U(N)$ gauge transformations
\be
\delta V_a = D_a \chi,  \hsp{2cm}
\delta X^i = i [\chi, X^i].
\ee
One should note however that the presence of $U(N)$ covariant pullbacks has consequences on the
invariance under NS-NS and R-R gauge transformations. Let us look for example at the variation
$\delta C_\mn = 2 \part_{[\mu} \Lambda_{\nu]}$ of the term given in (\ref{pullback}). Naively filling
in the variation in the pullback yields:
\be
\delta \{ P[C_2]\}  = 2\{ P[\part \Lambda_1] \}
          = 2\{ \part_\mu\Lambda_\nu \ D_{[a} X^\mu D_{b]}X^\nu \}.
          \label{derivtot}
\ee
In the Abelian limit this gauge variation is a total derivative, such that the
$\Lambda_1$ gauge invariance is assured in the D1-brane Chern-Simons action (for world volumes
without boundaries). In the non-Abelian case however the variation is not a total derivative
$$ 2\{ \part_\mu\Lambda_\nu \ D_{[a} X^\mu D_{b]}X^\nu \}
          \neq 2\part_{[a} \Bigl\{\Lambda_\nu \  D_{b]}X^\nu + ... \Bigr\}$$
and not even D1-branes with topologically trivial world volumes are described by a gauge
invariant action. We need to think more carefully how a background field gauge transformation
should be defined in the non-Abelian action.

The most important modification to the action (\ref{GHTac}),
with interesting physical implications, was found by Myers \cite{Myers}
and Taylor and Van Raamsdonk \cite{TvR}, when new dielectric couplings
to higher order background field potentials were shown to arise as a consequence of
T-duality in non-Abelian actions. Myers found that
the full T-duality invariant form of the Chern-Simons action
is given by:
\bea
S_{{\rm D}p}\ =\ T_p \int \Bigl\{ P[e^{(\incl_X \incl_X)} (C e^B) ] \  e^F \Bigr\},
\label{myersactie}
\eea
where $(\incl_X C)_{\mu_1\dots \mu_n}$ denotes the interior product
$X^{\mu_0} C_{\mu_0\dots \mu_n}$.
These terms induce dipole and higher moment couplings to R-R fields with rank higher
than $p+1$, which give rise, in particular, to the many
non-Abelian solutions of D$p$-branes expanding into fuzzy surfaces that have been constructed
in the literature (see for instance \cite{Myers, TV}).

Although these new terms are indeed necessary for the consistency of the T-duality
transformations, they make the issue of the gauge invariance of the non-Abelian action
even more unclear, because a pullback of a variational parameter of the form
$\{P[ (\incl_X \incl_X) \part \Lambda_{n+1} ]\}$ is by no means a total derivative.

{}From all this it is clear that the gauge invariance of the non-Abelian
action for D-branes requires further study. Nevertheless, there has been remarkably
few concern about this issue in the literature (see however \cite{Schwarz}). The first
serious but quite involved attempt to show that the Chern-Simons action (\ref{myersactie}) can
be written in
a form which is invariant under $\delta C_p = p\,\part \Lambda_{p-1}$ was made in
\cite{Ciocarlie}. By expanding the R-R background fields in their non-Abelian Taylor expansion
(\ref{taylorexp}), and integrating each term by parts, it was proven \cite{Ciocarlie} that the
Chern-Simons action could be written as an infinite series of terms involving only the R-R
field strengths. It is remarkable however that the resulting series cannot be interpreted
as a Taylor series, or as a pullback of $F_{p+1} (X^i)$. In spite of these non-trivial results,
\cite{Ciocarlie} can still not be the complete story, because it hardly looks at the NS-NS gauge
transformations
\be
\delta B_\mn = 2\part_{[\mu} \Sigma_{\nu]},
\label{AbNS}
\ee
neither does it address the massive gauge transformations of Romans' theory (which originally
concerned the authors of \cite{BR, GHT}).

The aim of this letter is twofold. First, by carefully defining what is meant by a background
gauge transformation in the non-Abelian world volume theory, we  will present an elegant method
to prove the invariance of the world volume effective action under the R-R and NS-NS gauge
transformations. Secondly, we will consider the most general case of a massive Type IIA
supergravity background, which will require dealing as well with the massive gauge
transformations of Romans' theory \cite{BR,GHT} for the dielectric case. In this respect
it is a priori not clear whether and how the action (\ref{myersactie}) will have to be
modified to the case of coincident D-branes in backgrounds of massive Type IIA supergravity.
The original derivation \cite{Myers} of the action (\ref{myersactie}) started with
$N$ coincident D9-branes and generated the action for coinciding lower-dimensional D-branes
through a chain of (massless) T-duality transformations, with more and more non-Abelian couplings
being generated as the number of transverse dimensions grew bigger. Since the D8-branes live in
massive IIA supergravity the appropriate setting would be to use the massive T-duality rules
between Romans' theory and Type IIB supergravity \cite{BRGPT, BR}, at least in the T-duality
between the D9- and the D8-branes and between the D8- and the D7-branes.

We will generalise the action (\ref{myersactie}) to include D-branes living in
massive Type IIA in two ways. First of all we will show that certain (mass dependent) world volume
couplings have to be added to the action to achieve invariance under massive gauge
transformations. Remarkably, we will find that no massive dielectric couplings are necessary
and that the only massive world volume couplings
are the ones already given in \cite{GHT}. Secondly  we will start from the action
for $N$ D9-branes and rederive the equivalent to (\ref{myersactie}) in a massive background
using massive T-duality. We will show that the extra massive terms previously added by demanding
gauge invariance are precisely those generated by the massive T-duality transformations, and that
no massive dielectric terms appear by this procedure either.

The organisation of this letter is as follows. In section 2 we discuss the gauge invariance
of the ``pre-dielectric'' action (\ref{GHTac}), taking into account the contribution of
the $U(N)$ covariant pullbacks and the fact that the background fields are functionals of the
non-Abelian scalars $X^\mu$. This will clarify how to define background gauge transformations
in the non-Abelian world volume theory. We show as well how this action has to be modified
to achieve invariance under massive gauge transformations.
In section 3 we generalise these
results to the full non-Abelian action including dielectric couplings and in section 4 we
show that these results agree with the ones obtained applying massive T-duality. Finally we
summarise our conclusions in section 5. Our conventions
for the gauge transformations of the background
fields and the Abelian actions can be found in appendix A.

\sect{Gauge invariance of the ``pre-dielectric'' action}

As mentioned in the introduction, in the non-Abelian case the gauge variation of the
pullback of a background field can no longer be defined as the pullback of the
gauge variation, simply because due to the covariant derivatives used in the pullback,
a naive pullback of the variational parameter can not be written as a total derivative,
and hence does not lead to a gauge invariant action. Moreover, adding more terms to the
action via a Noether procedure in order to obtain a
total derivative gauge variation would violate $U(N)$ covariance.

In order to have an action invariant under the background gauge transformations, we need
to fulfil two conditions. First, it must be possible to write the variation as a total
derivative, and second, the variation has to be a scalar under $U(N)$ gauge transformations.
We will propose a definition for a background gauge transformation in the non-Abelian world
volume theory that satisfies these conditions, and relates to the usual gauge transformations
in the Abelian case.

We define the variation of the pullback of a R-R field $C_p$ under the background
gauge transformation $\delta C_{p} =p\ \part \Lambda_{p-1}$ as:
\bea
\delta  P[C_{p}] \Omega \ \equiv \   p\  D P[\Lambda_{p-1}] \Omega
                         \  =  \  p\  D_{[a_1} (\Lambda_{\mu_2 ... \mu_{p}}
                                      D_{a_2}X^{\mu_2}... D_{a_p]}X^{\mu_p} )\Omega,
\label{defvar}
\eea
where $\Omega$ is any combination of world volume or pullbacked background fields and where
it is understood that all $U(N)$-valued objects appear symmetrises (though not in a trace).
In particular for the simplest case with $\Omega= 1$ we find that 
\bea
\delta  P[C_p]
   &=&  \  p\  D_{[a_1} (\Lambda_{\mu_2 ... \mu_{p}}
                                  D_{a_2}X^{\mu_2}... D_{a_p]}X^{\mu_p} )\nn
   &=& p\  \part_\nu \Lambda_{\mu_1 ...\mu_{p-1}}
                    D_{[b}X^{\nu}D_{a_1}X^{\mu_1} ... D_{a_{p-1}]}X^{\mu_{p-1}} \nn
   && \hspace{2cm}
        + \ p(p-1) \Lambda_{\mu_1...\mu_{p-1}} D_{[b}D_{a_1}X^{\mu_1}... D_{a_{p-1}]}X^{\mu_{p-1}} \nn
   &=& p\  P[\part \Lambda_{p-1}] \
           + \ \tfrac{i}{2} p (p-1)\ \Lambda_{\mu_1...\mu_{p-1}} [F_{[ba_1}, X^{\mu_1}]
                    D_{a_2}X^{\mu_2}... D_{a_{p-1}]}X^{\mu_{p-1}}. \nonumber
\eea
With this definition we see that the variation is not just the pullback of the gauge parameter,
but contains as well a non-Abelian correction term proportional to $[F, X]$, since the covariant 
derivative  $D_{a_1}$ not only acts on the background gauge parameter $\Lambda_{p-1}$, but also on 
the covariant derivatives in the pullback. For the Abelian case, the correction term disappears
and we recover the well-known gauge transformation for Abelian D-brane actions. Furthermore once we
consider terms in the action and trace over all $U(N)$ indices in the symmetrised trace prescription
the variation is in fact a total derivative:
\be
\delta \{ P[C_{p}] \} = p\ \{ D P[\Lambda_{p-1}]\}
                      = p \part \{ P[\Lambda_{p-1}]\}.
\label{deltaC}
\ee
Analogously, it is straightforward to derive from (\ref{defvar}) the variation of terms of the 
following forms:
\bea
\delta \{ P[C_{p}] \} &=& \Bigl\{ \sum_{n=0}^{[\tfrac{p-1}{2}]} \tfrac{p!}{2^n n! (p-2n-1)!}
                                D P[\Lambda_{p-2n-1}] P[B^n]  \Bigr\} .
\label{deltaCgen}\\
\delta \{ P[C_{p}] F^k \} &=& \Bigl\{ \sum_{n=0}^{[\tfrac{p-1}{2}]} \tfrac{p!}{2^n n! (p-2n-1)!}
                                D P[\Lambda_{p-2n-1}] P[B^n] F^k \Bigr\}
\label{deltaCF}\\
\delta \{ P[C_{p} B^k] \} &=& \Bigl\{ \sum_{n=0}^{[\tfrac{p-1}{2}]} \tfrac{p!}{2^n n! (p-2n-1)!}
                                D P[\Lambda_{p-2n-1}] P[B^{n+k}]  \Bigr\}
\label{deltaCB}
\eea

Similarly the non-Abelian generalisation of the NS-NS gauge transformation (\ref{AbNS}) is defined as:
\be
\delta P[B] = 2 D P[\Sigma] .
\label{nAbNS}
\ee
The Born-Infeld field should then transform as well as $\delta V=-P[\Sigma]$ , which is the
non-Abelian generalisation of the Abelian transformation
$\delta V_a=-\Sigma_\mu \partial_a X^\mu$.
It is clear that in this way the non-Abelian field strength $\cF = F + P[B]$ is indeed
invariant.

With these definitions, the computation of the gauge transformations of the action
\bea
\cL_{Dp} &=& (-)^{[(p+1)/2]} \left\{ P\left[\tfrac{1}{(p+1)!}C_{(p+1)}
                \ -\ \tfrac{1}{2(p-1)!} C_{p-1}\cF
                \ +\ \tfrac{1}{8(p-3)!} C_{p-3}\cF^2
                \right.  \right. \nn
&& \hspace*{1.5cm}
                \ -\ \tfrac{1}{48(p-5)!} C_{p-5}\cF^3
                \ +\ \tfrac{1}{384(p-7)!} C_{p-7}\cF^4
                \ -\ \tfrac{1}{3840(p-9)!} C_{p-9}\cF^5 \nn
&& \hspace*{1.5cm} \left. \left.
                \ +\ (-2)^{(p+2)/2}\ \tfrac{(p+1)!!}{(p+2)!}m \ \omega_{p+1} \right] \right\},
\label{mGHTac}
\eea
is straightforward, since it formally reduces to the Abelian case. Note that an extra
Chern-Simons-like term
\bea
\omega_{2n+1} = \sum_{k=0}^n \ \tfrac{i^k}{2^k (n+k+1)} \tfrac{(n+1)!}{k! (n-k)!} \
                             V (\part V)^{n-k} [V, V]^k,
\label{omega}
\eea
has to be added to the action of the even D-branes \cite{GHT}, in order to assure the invariance
under the
massive gauge transformations (\ref{gaugevar}) of massive Type IIA supergravity, which in the action
act on the pullback of the fields as
\be
\delta \{P[C_{2p+1}]\}  = - (2p+1)!!\  m\ \{ P[\Sigma \  B^p]\}.
\label{mC}
\ee
These Chern-Simons terms are constructed in such a way that they transform under the Yang-Mills gauge
transformations as a total derivative, and under the $\Sigma$ transformations as
\be
\delta \omega_{2n+1} = - \tfrac{n+1}{2^n}\ \Sigma F^n,
\ee
and thus cancel the massive gauge transformation of the R-R background fields. They are the
non-Abelian generalisation of the $V(\part V)^n$ terms in \cite{BR}.

So far we have rederived the results of \cite{GHT} on the gauge invariance of non-Abelian
Chern-Simons actions, taking into account explicitly the $U(N)$ covariant
pullbacks and the fact that the background fields are functionals of the non-Abelian
coordinates $X^\mu$. As we have seen this forces a precise definition for what we mean
by gauge variation of a non-Abelian pullback. A consistency check of our
definitions (\ref{deltaC})-(\ref{nAbNS}) and (\ref{mC}) is that the variation of
the pullback of a R-R $p$-form should be T-dual to the variation of the pullback of a
R-R $(p-1)$-form field. In the next section we will check this and see that in this manner we
can find a natural way to also prove the gauge invariance of the dielectric terms.

\sect{Gauge invariance of the dielectric couplings}

In this section we show that the gauge transformation of the pullback of a R-R field
$C_p$, as defined in (\ref{defvar}), is consistently mapped under T-duality into the
gauge transformation of the pullback of the T-dual of $C_p$. To show this let us
define a R-R field $\tC_p$, being related to $C_p$ via a gauge transformation\footnote{In
this section a hatted index indicates that it runs from 0 to 9, including the T-duality
direction, which is denoted by $x$. Unhatted indices exclude the T-duality direction. Similarly in
the world volume indices we denote the T-duality direction by $\sigma$ and the other directions by
$a$.}
$\tC_{\hmu_1 ... \hmu_p} = C_{\hmu_1 ... \hmu_p} + p\ \part_{[\hmu_1} \Lambda_{\hmu_2 ...\hmu_p]}$.
We then have on the one hand by definition (\ref{defvar}) that
\be
\{ \tC_{a_1 ... a_{p-1} \sigma} \} = \{ C_{a_1 ... a_{p-1} \sigma} \}
        \ +\ p\ \part_{[a_1} \{ \Lambda_{a_2 ... a_{p-1} \sigma]}  \},
\ee
while on the other hand we know \cite{Myers} that applying T-duality on $\tC_p$ we get
(for simplicity we truncate for now to the ``diagonal approximation'' $g_{\mu \ux} = B_\hmn = 0$)
\bea
&&  \hspace{-1cm}
\{ \tC_{\hmu_1 ... \hmu_p} D_{[a_1} X^{\hmu_1} ... \
                           D_{a_{p-1}} X^{\hmu_{p-1}} D_{\sigma]} X^{\hmu_p} \} \rightarrow \\[.2cm]
&&\rightarrow \{ \tC_{\mu_1 ... \mu_{p-1}} D_{[a_1} X^{\mu_1} ... \
                                   D_{a_{p-1}]} X^{\mu_{p-1}}
     \ +\ i \tC_{\mu_1 ... \mu_p x} D_{[a_1} X^{\mu_1} ... \
                                      D_{a_{p-1}]} X^{\mu_{p-1}} [X^x, X^{\mu_p}] \}
\nonumber \\[.3cm]
&&= \{ C_{\mu_1 ... \mu_{p-1}} D_{[a_1} X^{\mu_1} ... \
                                   D_{a_{p-1}]} X^{\mu_{p-1}}
      \ +\ i C_{\mu_1 ... \mu_p x} D_{[a_1} X^{\mu_1} ... \
                                      D_{a_{p-1}]} X^{\mu_{p-1}} [X^x, X^{\mu_p}] \} \nn
&& \hsp{.5cm}
     +\ \part_{[a_1}\{ (p-1)\ \Lambda_{\mu_2 ... \mu_{p-1}}D_{a_2} X^{\mu_2} ... \
                                   D_{a_{p-1}]} X^{\mu_{p-1}}      \nn
&&  \hsp{3cm}
                      + (p+1)\  i \ \Lambda_{\mu_2 ... \mu_p x}D_{a_2} X^{\mu_2} ... \
                                   D_{a_{p-1}]} X^{\mu_{p-1}}[X^x, X^{\mu_p}]   \},  \nonumber
\eea
where we used that $\tC_{p-1}$ and $\tC_{p+1}$ are related to, respectively, $C_{p-1}$ and $C_{p+1}$
by the same type of background gauge transformation that relates $\tC_p$ to $C_p$. We then find that
the pullback of the gauge parameter transforms under T-duality as
\be
p \ \{ D_{[a_1} \Lambda_{a_2 ... a_{p-1} \sigma]} \} \rightarrow
 \{ (p-1)\ D_{[a_1} \Lambda_{a_2 ... a_{p-1}]}
\ +\ (p+1) \ i \ D_{[a_1} \Lambda_{a_2 ... a_{p-1}] \mu_p x} [X^x, X^{\mu_p}]   \}.
\ee
In other words, the variation of the pullback of a R-R $p$-form potential
goes under T-duality to the
variation of the pullback of a R-R $(p-1)$-form potential
plus the variation of the pullback of the first dielectric coupling term:
\be
\delta \{ P[C_p] \} \ \rightarrow \
\delta \{ P[C_{p-1}] \}\  + \ i \ \delta \{ P[ (\incl_X \incl_X) C_{p+1}] \},
\ee
if we define:
\bea \delta \{ P[ (\incl_X\incl_X) C_{p+1}] \}
&\equiv& (p+1) \part \{ P[ (\incl_X \incl_X) \Lambda_{p}]  \} \\ [.3cm]
&=& (p+1) \{ P[ (\incl_X\incl_X) \part \Lambda ]
  \ + \ (p-2) \Lambda_{ \hmu \hnu\hrho [a_3 ... a_{p-1}} [ F_{a_1 a_2]}, X^\hrho] [X^\hnu, X^\hmu] \nn
&& \hsp{1cm}
  \ + \ 2 \Lambda_{\hmu \hnu[a_2 ... a_{p-1} } [D_{a_1]}X^\hnu, X^\hmu]) \}. \nonumber
\eea
The derivation with the full T-duality rules (beyond the diagonal approximation) is straightforward
and not very enlightening, so we rather concentrate on the generalisation of the variation
(\ref{deltaCgen}) for dielectric couplings,
which can be derived in a similar way. Under general R-R gauge
transformations (\ref{gaugevar}) the dielectric terms vary as
\bea
\delta_\Lambda \{ P[(\incl_X \incl_X) C_{p}] \}
        &=& \sum_{n=0}^{[(p-1)/2]} \Bigl\{
              \tfrac{(p-2)!(p-2n)}{2^n n! (p-2n-2)!} D P[\incl_X \incl_X\Lambda_{p-2n-1}] P[B^n] \nn
      && \hsp{1cm}
            +\ \tfrac{(p-2)!(p-2n)}{2^{n-2} (n-1)! (p-2n-1)!} D P [\incl_X\Lambda_{p-2n-1}]
                                     P[(\incl_X B)B^{n-1}]  \label{Lambdadielec}
                                                             \\[.2cm]
       && \hsp{1cm}
            +\ \tfrac{(p-2)!}{2^{n-2} (n-2)! (p-2n-1)!} D P [\Lambda_{p-2n-1}]
                                     P[(\incl_X B)^2 B^{n-2}]  \nn
        && \hsp{1cm}
            +\ \tfrac{(p-2)!}{2^{n-1} (n-1)! (p-2n-1)!} D P [\Lambda_{p-2n-1}]
                                     P[(\incl_X \incl_X B) B^{n-1}] \nonumber
                            \Bigr\}.
\eea
Here the coefficients in front of each term are the different weights that arise when
the inclusion factor $(\ii)$
acts on the various background fields.

Similarly, under massive gauge transformations, the dielectric terms
transform as
\be
\delta_\Sigma \{ P[(\incl_X \incl_X) C_{p}] \} = - p!! \ m \ \{ P[ (\ii) (\Sigma B^{(p-1)/2}) ] \},
\label{mdielec}
\ee
while the variation under the NS-NS gauge transformation is given by
\bea
 \delta_\Sigma \{ P[\incl_X B]\} = \{ 2 DP[ \incl_X \Sigma] \},
\hsp{1.5cm}
 \delta_\Sigma \{ P[(\incl_X \incl_X) B]\} = \{  (\ii) \part \Sigma \}.
\label{sigmadielec}
\eea

As an example let us now look at the gauge transformations of the non-Abelian action for
D6-branes, being this the simplest non-trivial case in which both dielectric couplings and
massive gauge transformations are present. We will leave the invariance of the general D$p$-brane
action to the reader.

The dielectric part in the non-Abelian Chern-Simons action describing a system of coincident
D6-branes in a massive Type IIA background can be written as (see Appendix A)
\bea
\cL_{D6} \sim  - \left\{ \sum_{n=0}^{3} \tfrac{(-)^n}{2^n n!(7-2n)!}
                                P\Bigl[ (\incl_X \incl_X) \cA_{9-2n} \Bigr] F^n \right\},
\label{nonAbaction}
\eea
where the $p$-forms $\cA_{p}$ are defined in (\ref{cA}).
This form is very convenient to show the R-R gauge invariance, because
it is obvious from the Abelian case that each $\cA_{p}$ is
invariant under the R-R and massive gauge transformations (\ref{gaugevar}). Therefore,
in this form the invariance of the action (\ref{nonAbaction})
under the transformations (\ref{Lambdadielec}) and (\ref{mdielec}) is straightforward. It is
also clear
that besides the massive terms (\ref{omega}), introduced in \cite{GHT}, no other dielectric
mass terms are
needed to assure gauge invariance.

The invariance under the NS-NS transformations (\ref{sigmadielec}) is however more subtle,
due to the fact that $(\incl_X\incl_X)$ acts on $B$ but it does not act on $F$, so that they
do not combine
in an obvious way into the interior product of the gauge invariant field strength
$\cF$. In order to show the invariance under these transformations let us look at those
terms with a given $C_p$, for
example the dielectric terms that couple to $C_7$
\be
\cL_{D6} \sim \Bigl\{ P[ -21 (\incl_X \incl_X)C_7 F \ - \  36 (\incl_X \incl_X) (C_7 B)] \Bigr\}.
\ee
Taking into account that the interior products
$(\incl_X \incl_X)$ in the first
term only act on $C_7$, while in the second term they act both on $C_7$ and on $B$,
we have, explicitly:
\be
\cL_{D6} \sim \Bigl\{ P[ -21 (\incl_X \incl_X C_7) F \ - \  21 (\incl_X \incl_X  C_7)B
                         \ -\ 14 (\incl_X C_7)(\incl_X B)\ -\ C_7 (\incl_X \incl_X B) ]\Bigr\},
\ee
so we see that the first two terms can be combined into the NS-NS invariant field
strength,
whereas the terms where $B$ is contracted with one or two $\incl_X$ cannot. The NS-NS
gauge variation of the latter is given by
\be
\delta_\Sigma \cL_{D6} \sim \Bigl\{
          -28 P[ (\incl_X C_7)]DP[(\incl_X \Sigma)]\
          -\ 2 P[C_7] (\incl_X \incl_X \part \Sigma) ]\Bigr\},
\label{zeroterms}
\ee
and can not be canceled by any other term in the action. The only field that also transforms
under NS-NS transformations is the Born-Infeld vector $V_a$, which is however a world volume
field and cannot be contracted with $\incl_X$. Actually, these contractions of the gauge
parameter with the transverse scalars are zero, due to reasons inherent to the
construction of the action. Recall that the action for non-Abelian D$p$-branes is
derived from the action for coincident D9-branes using T-duality \cite{Myers}, so that
the directions in which the T-dualities are performed have to be isometric. In the
T-dualised action these isometry directions correspond to the transverse directions
$X^i$, which through the T-duality mapping $V_i\rightarrow X^i$ inherit the gauge
transformation of the $i$-th component of the Born-Infeld field:
$\delta_\Sigma X^i=-\Sigma^i$. Since these directions are isometric, the contractions
of $\Sigma$ with the transverse scalars must vanish, and in this way the gauge invariance
is guaranteed. Strictly speaking this kind of terms are already zero in the Abelian case.
However in that case if we demand the action to be invariant under diffeomorphisms of
the background we recover a fully invariant action.
Yet, in the non-Abelian case there is no clear notion of general coordinate transformations (see
for example \cite{DS, DSW}), and it is not clear how the resulting isometries can be removed.
The fact that the terms in (\ref{zeroterms}) vanish is therefore a manifestation of the lack of
diffeomorphism invariance of the action (\ref{D6}). We expect that in a fully diffeomorphism
invariant formulation extra terms will be generated whose variation will cancel the terms
in (\ref{zeroterms}).

\sect{The non-Abelian CS action and massive T-duality}

In this section we derive the Chern-Simons action for coincident Type IIA D$p$-branes in
a massive background, applying the same method of \cite{Myers}. We start with a system of
$N$ D9-branes and use the massive T-duality rules between the massive Type IIA and Type IIB
theories to generate the action for lower-dimensional branes. We will show that the form
of the action is consistent with the one obtained in the previous sections from gauge invariance.
In particular we will show that, besides the mass terms (\ref{omega}) presented in \cite{GHT} no
extra dielectric mass terms are required.

The massive T-duality rules for general R-R potentials are given by \cite{BRGPT, EL}
\bea
C_{2k}|_x &=& -C_{2k-1} \ +\ (2k-1) \tfrac{g|_x}{g_{xx}} C_{2k-1}|_x
             \ - \ m x \tfrac{(2k-1)!}{2^{k-1}(k-1)!} \tfrac{g|_x}{g_{xx}} B^{k-1} \nn
C_{2k} &=& C_{2k+1}|_x \ -\ 2kC_{2k-1}B|_x
            \ -\ 2k(2k-1)C_{2k-1}|_x B|_x \tfrac{g|_{x}}{g_{xx}} \nonumber \\
       && \hsp{.5cm}  - \ \tfrac{(2k)!}{2^k k!} mx B^k
         \ -\ \tfrac{(2k)!}{2^{k-1}(k-1)!} mx \ \tfrac{g|_x}{g_{xx}} B^{k-1}B|_x
\label{massiveT} \\[.2cm]
C_{2k+1}|_x &=& C_{2k}\ +\ 2k \tfrac{g|_x}{g_{xx}} C_{2k}|_x \ + \ \tfrac{(2k)!}{2^k k!} mx B^k
               \ +\ \tfrac{(2k)!}{2^{k-1}(k-1)!} mx \tfrac{g|_x}{g_{xx}} B^{k-1}B|_x    \nn
C_{2k+1} &=& -C_{2k+2}|_x \ +\ (2k+1)C_{2k}B|_x \ + \ (2k+1)2k  \tfrac{g|_x}{g_{xx}} C_{2k}|_x B|_x
\nonumber
\eea
where $x$ denotes the T-duality direction and $.|_x$ means that the last space-time index is $x$.

The action for $N$ D9-branes can be expressed in terms of the $p$-forms $\cA_p$ of (\ref{cA}) as
\be
\cL_{D9} = - \Bigl\{ \sum_{n=0}^{5} \tfrac{(-)^n}{2^n n! (10-2n)!} \cA_{10-2n} F^n \Bigr\}.
\ee
Taking into account that the massive T-duality rules (\ref{massiveT}) acting on (\ref{cA})
give simply
\bea
\cA_{2q +3}|_x \leftrightarrows \cA_{2q+2}, \hsp{1cm}
\cA_{2q+2}|_x \leftrightarrows - \cA_{2q+1}, \hsp{1cm}
\cA_1|_x \leftrightarrows C_0 + mX^x,
\label{TdualcA}
\eea
and that the Born-Infeld vector and the covariant pullback transform under T-duality as \cite{Myers}
\be
\begin{array}{lll}
V_\sigma \rightarrow X^x  \hsp{1.5cm}
   &  D_a X^{\mu} \rightarrow D_a X^{\mu},  \hsp{1.5cm}
   &  D_\sigma X^{\mu} \rightarrow i[X^x,X^{\mu}],  \\[.2cm]
F_{a\sigma} \rightarrow D_a X^x \hsp{1.5cm}
   &  D_a X^x \rightarrow 0,
   &  D_\sigma X^x \rightarrow 1,
        \end{array}
\label{TdualDX}
\ee
we obtain for a system of $N$ coincident D$p$-branes
\bea
\cL_{Dp} &=& (-)^{[\tfrac{p+1}{2}]} \left\{ \sum_{n=0}^{[\tfrac{p+1}{2}]} \tfrac{(-)^n}{(2n)!!(p+1-2n)!}
       P \Bigl[  \cA_{(p+1-2n)}
               - i (\incl_X \incl_X) \cA_{p+3-2n}
               - \tfrac{1}{2!}(\incl_X\incl_X)^2 \cA_{p+5-2n}
                                                  \right. \nonumber\\
&&\left.
               + \tfrac{i}{3!}(\incl_X\incl_X)^3 \cA_{p+7-2n}
               + \tfrac{1}{4!}(\incl_X\incl_X)^4\cA_{p+9-2n} \Bigr] F^n
               +\ (-2)^{\tfrac{p+2}{2}}\ \tfrac{(p+1)!!}{(p+2)!}m \ \omega_{p+1}
                                                        \vphantom{\sum_{n=0}^{[(p+1)/2]}}  \right\}.
\label{generalDp}
\eea
It is clear from the T-duality rules (\ref{TdualcA}) and (\ref{TdualDX}) that the only massive
terms that appear in the action come from the $C_0 F^{p/2} F|_\sigma$ terms, which after partial
integration \cite{BR} lead to the $\omega_{p+1}$. This agrees with the conclusions from section 3,
where it was found that the resulting action was invariant under massive gauge transformations
without the need for extra dielectric massive terms.

\sect{Conclusions}

We have presented an elegant way to demonstrate the invariance of the non-Abelian Chern-Simons
action for D-branes under the gauge transformations of the background fields. By
carefully defining
what precisely is meant with a background gauge variation in the world volume theory, we
have found
that the action is indeed invariant under R-R transformations (including
massive transformations) if
a Chern-Simons-like mass term involving the non-Abelian Born-Infeld vector is added. This extra
mass term appears only in the non-dielectric part of the action and its presence was already found
in \cite{GHT}. No extra dielectric mass terms are present. Our results are confirmed by the
construction of the non-Abelian Chern-Simons action for D-branes in massive backgrounds via
massive T-duality.

The invariance under the NS-NS transformations is more subtle, but it is guaranteed because
by (the T-duality) construction the transverse space directions must be isometric. As a
consequence of this the variation of the terms
where the NS-NS $B$-field is contracted with one or two inclusions $\incl_X$ vanishes. The fact
that these terms are zero is intimately related to the fact that the non-Abelian Chern-Simons
action is not invariant under general coordinate tranformations. We would
expect a fully diffeomorphism
invariant action to contribute with extra terms whose variation under NS-NS
gauge transformations would cancel the variations that in our case are set to zero by construction.
The structure of the latter could give
valuable hints in the construction of a diffeomorphism invariant Chern-Simons action for D-branes.

\vspace{1cm}

\noindent
{\bf Acknowledgements}\\
We wish to thank Mees de Roo, Martijn Eenink, Paul Koerber, Patrick Meessen and Alex Sevrin 
for the useful discussions.
B.J. is grateful to the Instituto de F\'{\i}sica Te\'orica of the U.A.M./C.S.I.C. in Madrid,
where part of this work has been done.
This work has been made possible with the aid of the  F.W.O.-Vlaanderen, to which J.A. and J.G.
are associated as Aspirant F.W.O. and B.J. as a Post-doctoral Fellow. J.A., J.G. and B.J. were
also partially supported by the European Commission R.T.N. program HPRN-CT-2000-00131, by the
F.W.O.-Vlaanderen project G0193.00N and by the Belgian Federal Office for Scientific, Technical
and Cultural Affairs through the Interuniversity Attraction Pole P5/27. The work of Y.L. has been
partially supported by CICYT grants BFM2000-0357 and BFM2003-00313 (Spain).

\appendix
\renewcommand{\theequation}{\Alph{section}.\arabic{equation}}

\sect{Conventions}

In our conventions the gauge transformations of the R-R background fields are given by:
\bea
\delta C_p = \sum_{n=0}^{[\tfrac{p-1}{2}]} \tfrac{p!}{(2n)!! (p-2n -1)!}
                              \part \Lambda_{p-2n-1} B^n
               \ -\ p!!\ m\ \Sigma B^{(p-1)/2},
\label{gaugevar}
\eea
where the square brackets in the summation indicate integer part for $p$ even, and the massive term
is only non-vanishing for the odd R-R potentials of the massive Type IIA theory.

The Abelian Chern-Simons action for D-branes, involving massive terms for the case of Romans' theory,
is given by \cite{BR, GHT}:
\bea
\cL_{Dq} =(-)^{[\tfrac{q+1}{2}]} \sum_{n=0}^{[\tfrac{q+1}{2}]}
                                       \tfrac{(-)^n}{2^n n!(q-2n+1)!} P[C_{q-2n+1}] \cF^n
               \ -\ (-)^{\tfrac{q}{2}}\    2^{\tfrac{q+2}{2}}\
                        \tfrac{(q+1)!!}{(q+2)!}m V (\part V)^{q/2}.
\label{Abaction}
\eea
Here $P[...]$ denotes the (Abelian) pullback and $\cF$ is defined as $\cF = F + P[B]$, with
$F= 2\part V$.

A useful form for the non-Abelian Chern-Simons action can be obtained introducing
the following $p$-forms
\be
\cA_p= \sum_{k=0}^{[\tfrac{p}{2}]}
       \tfrac{(-)^k p!}{2^k k!(p-2k)!}\ C_{p-2k} B^k,
\label{cA}
\ee
which contain
basically (up to an overall factor) the background field dependence of the pullback of
the Abelian D$(p-1)$-brane action (\ref{Abaction}). The non-Abelian action for, for example, a
D6-brane is then given by
\bea
\cL_{D6} =  - \left\{  \sum_{n=0}^{3} \tfrac{(-)^n}{2^n n!(7-2n)!} P[C_{7-2n}] \cF^n
                     \ + \ \tfrac{1}{24}\ m \ \omega_7
                     \ - \ \sum_{n=0}^{3} \tfrac{(-)^n}{2^n n!(7-2n)!}
                                P\Bigl[i (\incl_X \incl_X) \cA_{9-2n} \Bigr] F^n \right\},
\label{D6}
\eea
where now $F$ is the non-Abelian field strength $F= 2\part V + i [V, V]$ and the pullback is
taken with covariant derivatives. The first two terms constitute the ``pre-dielectric''
action
presented in \cite{Douglas, GHT}, while the third term contains the dielectric couplings of
\cite{Myers}. The fact that the dielectric couplings appear precisely through contractions of
the $p$-forms $\cA_p$ is crucial for the gauge invariance of the action. Lower dimensional
D-branes,
with higher rank dielectric couplings, will contain analogous structures.


\end{document}